\documentclass{article}
\usepackage{graphicx} 
\usepackage[a4paper, total={6in, 8in}]{geometry}
\usepackage{caption}
\usepackage{subcaption}
\title{Spatial clustering of temporal energy profiles with empirical orthogonal functions and max-p regionalization}
\author{Claire Halloran, Malcolm McCulloch\\
Department of Engineering Science, University of Oxford}
\date{\today}

\begin{document}

\maketitle
\begin{abstract}
    This paper presents a spatial clustering method to create regions with similar time-varying energy characteristics. This method combines empirical orthogonal functions (EOFs) for dimensionality reduction and max-p regionalization for spatial clustering. The proposed approach creates regions that each have a similar value of a spatially extensive attribute, such as available land area, population, or GDP, as well as similar weather-dependent temporal energy profiles, such as wind and solar generation potential or heating and cooling demand, within each region. We demonstrate this technique using hourly wind and solar generation potential in 2019 in Ireland and Britain. Solar generation clusters are best-defined at a smaller land area threshold compared to wind generation.
\end{abstract}

\section{Introduction}

This paper addresses the question: \emph{How can spatiotemporal energy data be used to create regions of similar size with similar time-varying energy characteristics?} We propose a spatial clustering method that combines empirical orthogonal functions (EOFs) for dimensionality reduction of temporal data and max-p regionalization for spatial clustering to address this problem.

As electricity systems become increasingly reliant on weather \cite{Staffell2018}, incorporating high spatial and temporal resolution data into energy systems modeling is necessary. One of the primary approaches for incorporating higher spatial and temporal resolution data into energy systems modeling while maintaining computational tractability is spatial clustering to create regions with representative energy characteristics. Two widely used methods for spatial clustering of energy data are k-means clustering \cite{Getman2015, horsch_role_2017, siala_impact_2019, scaramuzzino_integrated_2019, frysztacki_modeling_2020, Frysztacki2021, Frysztacki2022, jani_temporal_2022} and max-p regionalization \cite{Getman2015, siala_impact_2019, Fleischer2020, Fleischer2021}. These spatial clustering techniques are most commonly used on static spatial data, such as total annual energy demand or annual wind and solar potential \cite{horsch_role_2017, siala_impact_2019,scaramuzzino_integrated_2019, Fleischer2020, frysztacki_modeling_2020, Frysztacki2021, Fleischer2021}. 

These approaches based on static spatial data do not account for temporal differences in energy data within regions. To address this issue, a few works have proposed methods for spatial clustering of temporal energy data. Getman et al.~\cite{Getman2015} compare the use of k-means clustering and max-p regionalization of reduced temporal data to create regions with similar solar potential. More recently, Jani et al.~\cite{jani_temporal_2022} use k-means clustering to group areas with similar correlation coefficients for wind and solar to identify locations with complementary renewable potential, and Frysztacki et al.~\cite{Frysztacki2022} propose the use of hierarchical agglomerative clustering of wind and solar potential to create regions for energy systems planning. 

Using k-means clustering on temporal can result in regions of different sizes, for instance, areas with different annual renewable potential or annual demand. In this paper, we propose a spatial clustering method that combines EOFs for dimensionality reduction and max-p regionalization to create regions with similar time-varying energy characteristics where all regions have similar values of another spatially extensive attribute. This approach can be helpful in ensuring that all clustered regions represent a significant supply or demand potential. 

\section{Data and methods}
In this section, we discussion the data used to demonstrate this spatial clustering technique and detail the EOF analysis and max-p regionalization approach it uses.

\subsection{Data}
We demonstrate the proposed clustering technique using hourly onshore wind and solar generation potential in 2019 in Ireland and Britain at 0.3 degree by 0.3 degree spatial resolution (approximately 30 km by 30 km). We use the Python package atlite \cite{atlite} to calculate the generation potential for a Vestas V90 3MW wind turbine and a crystalline silicon photovoltaic based on reanalysis data from ERA5 \cite{ERA5}.

\subsection{EOF analysis}\label{sec: EOFs}

We use empirical orthogonal function (EOF) analysis for dimensionality reduction of wind and solar data prior to spatial clustering. EOF analysis is used in meteorology, oceanography, and climate science to identify key spatial patterns in spatiotemporal data. This technique involves calculating the eigenvalues and eigenvectors of a spatially weighted anomaly covariance matrix of a spatiotemporal field \cite{NCAR2013}. This procedure decomposes spatiotemporal data into a set of principal component (PC) time series and corresponding EOF spatial patterns, which capture the correlation between each the PC and the time series at each point in space.

We use the eofs python package \cite{Dawson2016} to perform EOF analysis on 1 year of hourly wind and solar generation data in Britain and Ireland in 2019. Because wind and solar generation potential are weather-driven, we expect that the first few PCs will capture most of the variance, as with weather processes \cite{NCAR2013}. We will limit our analysis to only the PCs and EOFs that account for greater than 1\% of the total variance. 

\subsection{Max-p regionalization}\label{sec: max-p regionalization}

The max-p regions problem seeks to maximize the number of spatially contiguous regions that meet a certain threshold value, in this case land area, while minimizing heterogeneity of features within a region, in this case the EOF correlation values \cite{Duque2012}. We apply max-p regionalization to the first 5 wind EOF values and the first 4 solar EOF values to create wind and solar regions. This means that each clustered region should have similar temporal variation in generation potential, as well as comparable total annual generation. 

We use the spopt Python library \cite{spopt2021, spopt2022} to perform max-p regionalization. Because the exact max-p regions problem is NP-hard, this library implements the heuristic approach introduced by Wei et al.~\cite{Wei2021}. For both the wind and solar EOFs, max-p regionalization is performed with area threshold values of 50,000, 70,000, and 100,000 km$^2$ to generate different sets of wind and solar regions with similar values of total annual generation. Rook contiguity is used for the contiguity constraint, so each spatial area within a region must share an edge with another area in that region.

\subsection{Cluster evaluation}

To evaluate the clusters resulting from this process, we use the Python library scikit-learn \cite{sklearn} to calculate the Silhouette Coefficient \cite{Rousseeuw1987} using cityblock distance based on the underlying hourly renewable potential. This metric compares the mean distance between a sample and all other points in the same cluster, with the mean distance between a sample and all points in the next nearest cluster, so scores closer to 1 indicate more well-defined clusters.

\section{Results}

In this section, we discuss the results of the EOF analysis and max-p regionalization clustering for the solar and wind data.

\subsection{EOF analysis}
\begin{figure}
    \centering
    \begin{subfigure}[b]{0.45\textwidth}
             \centering
             \includegraphics[width=\textwidth]{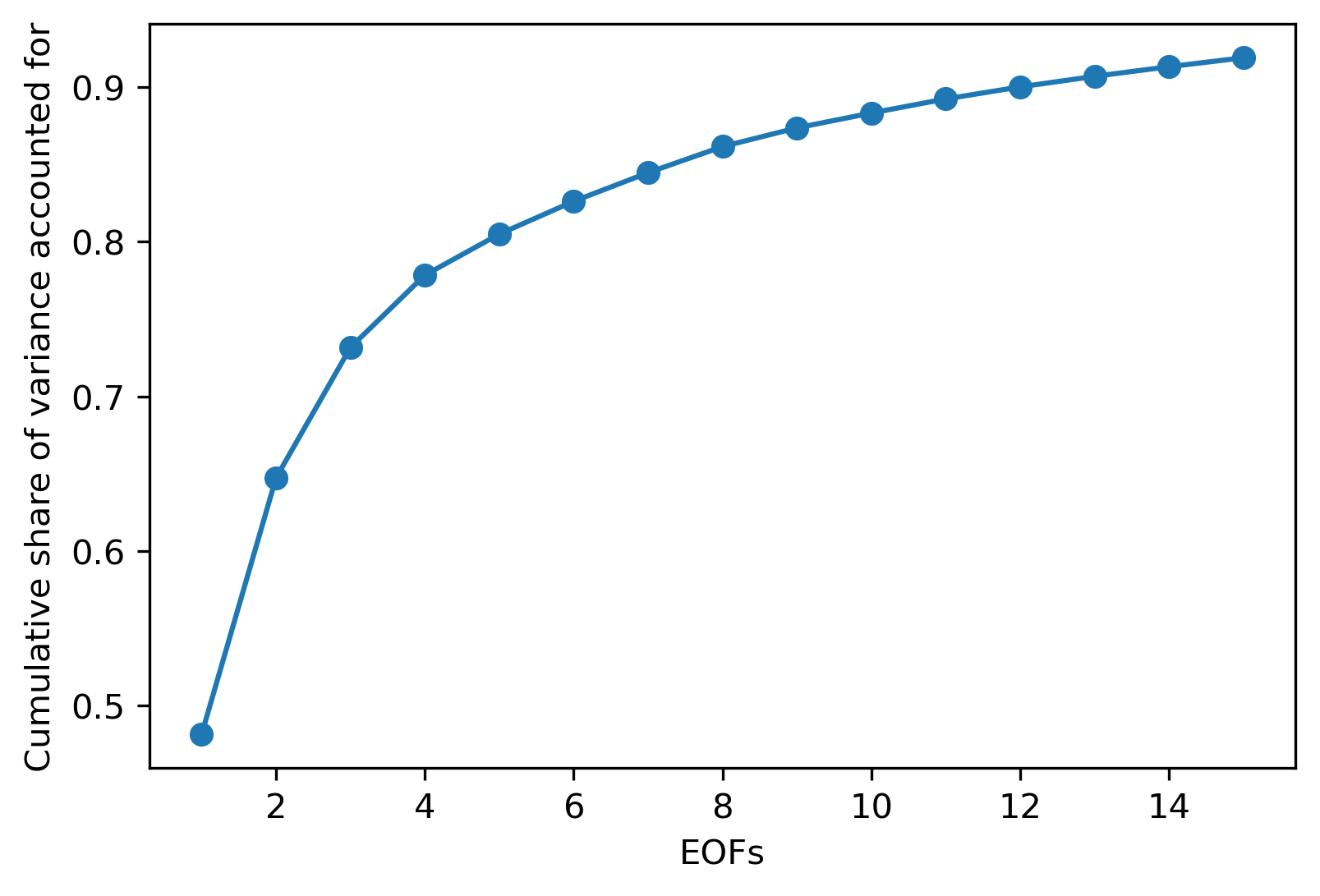}
             \caption{Wind variance vs. EOFs}
             \label{fig:wind variance}
         \end{subfigure}
         \hfill
    \begin{subfigure}[b]{0.45\textwidth}
             \centering
             \includegraphics[width=\textwidth]{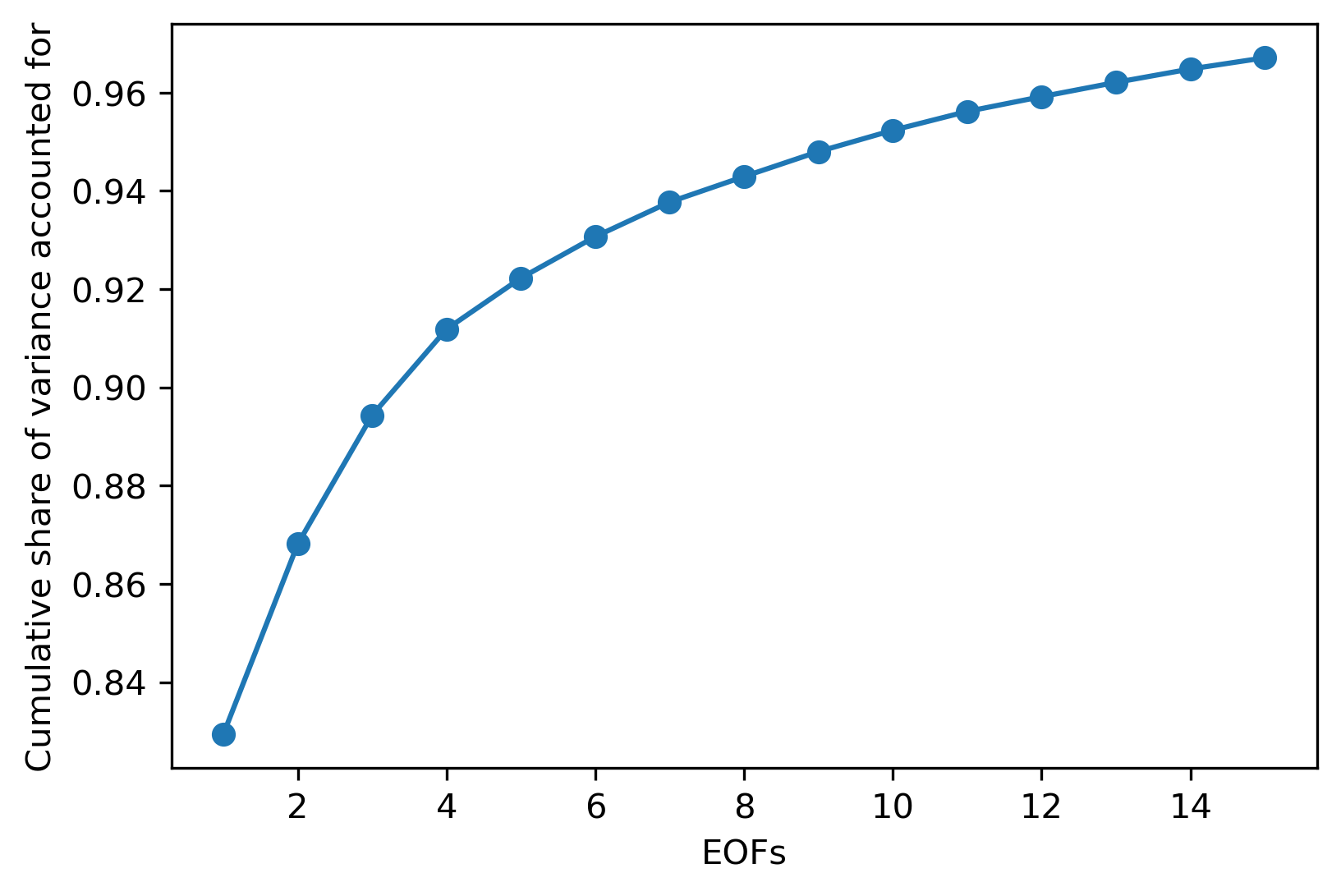}
             \caption{Solar variance vs. EOFs}
             \label{fig:solar variance}
    \end{subfigure}
    \caption{Cumulative share of wind and solar variance accounted for with each EOF and its corresponding PC.}
    \label{fig:variance}
\end{figure}

As expected for weather-driven quantities, the first few EOFs and PCs account for a large share of variance in the wind and solar generation potential. As shown in Figure \ref{fig:wind variance}, the first 5 EOFs account for 80\% of variance in onshore wind potential in Britain and Ireland, and the sixth EOF accounts for less than 1\% of variance. The first 4 EOFs account for 91\% of variance in onshore solar potential, shown in Figure \ref{fig:solar variance}, and the fifth EOF accounts for less than 1\% of variance. We therefore limit our analysis to the first 5 EOFs for wind and the first 4 EOFs for solar.

\begin{figure}
    \centering
    \begin{subfigure}[b]{0.3\textwidth}
             \centering
             \includegraphics[width=\textwidth]{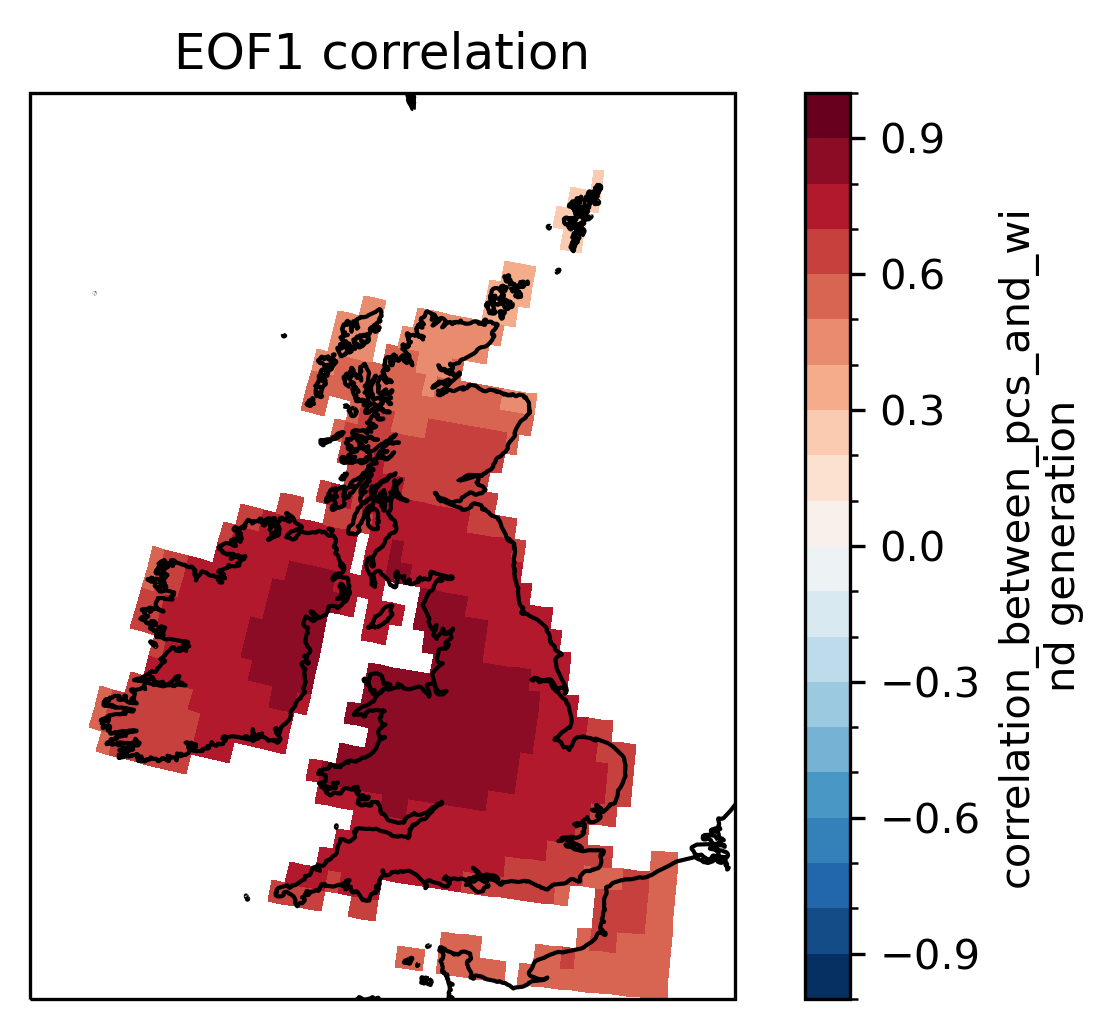}
             \caption{EOF 1}
             \label{fig:wind EOF1}
         \end{subfigure}
         \hfill
         \begin{subfigure}[b]{0.3\textwidth}
             \centering
             \includegraphics[width=\textwidth]{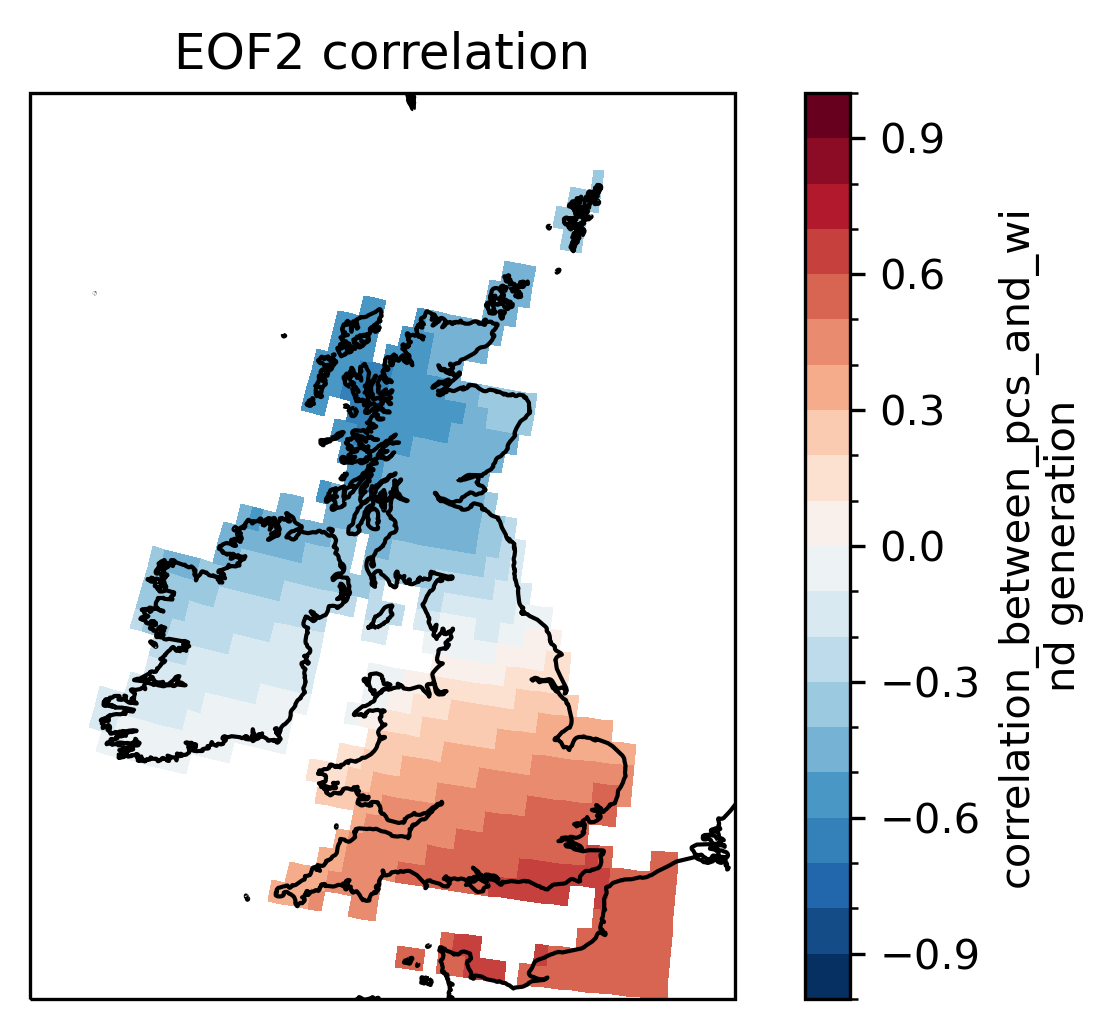}
            \caption{EOF 2}
             \label{fig:wind EOF2}
         \end{subfigure}
         \hfill
         \begin{subfigure}[b]{0.3\textwidth}
             \centering
             \includegraphics[width=\textwidth]{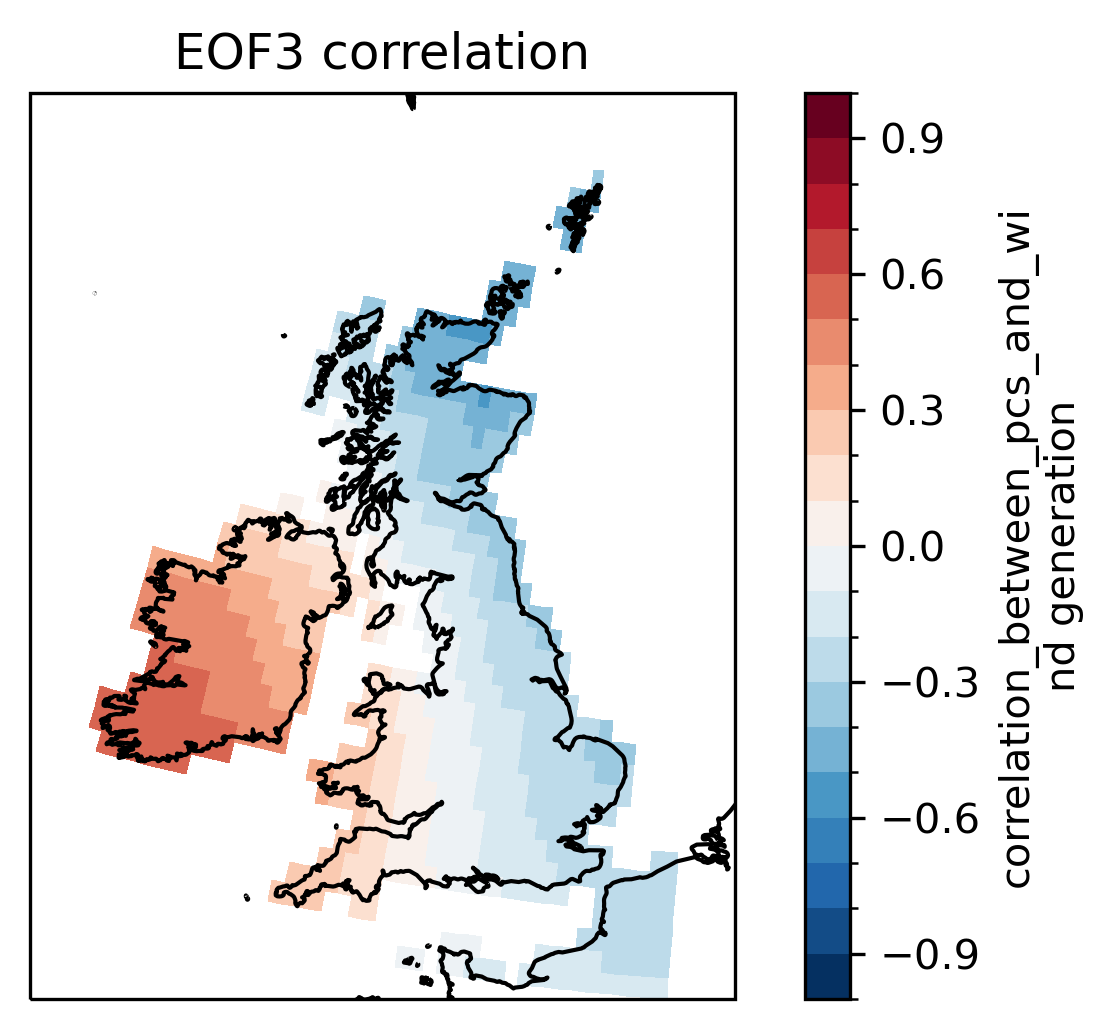}
            \caption{EOF 3}
             \label{fig:wind EOF3}
         \end{subfigure}
                  \begin{subfigure}[b]{0.3\textwidth}
             \centering
             \includegraphics[width=\textwidth]{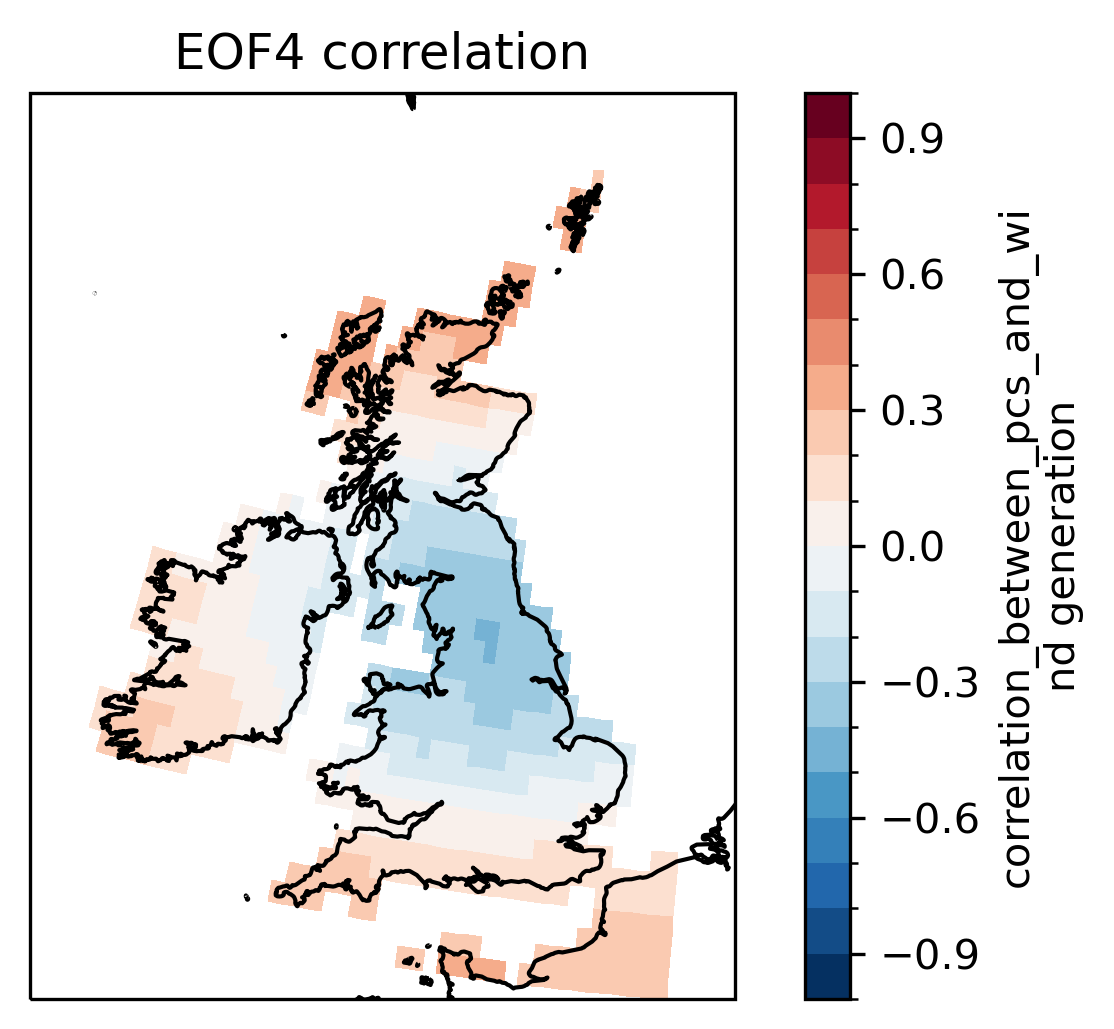}
             \caption{EOF 4}
             \label{fig:wind EOF4}
         \end{subfigure}
         \begin{subfigure}[b]{0.3\textwidth}
             \centering
             \includegraphics[width=\textwidth]{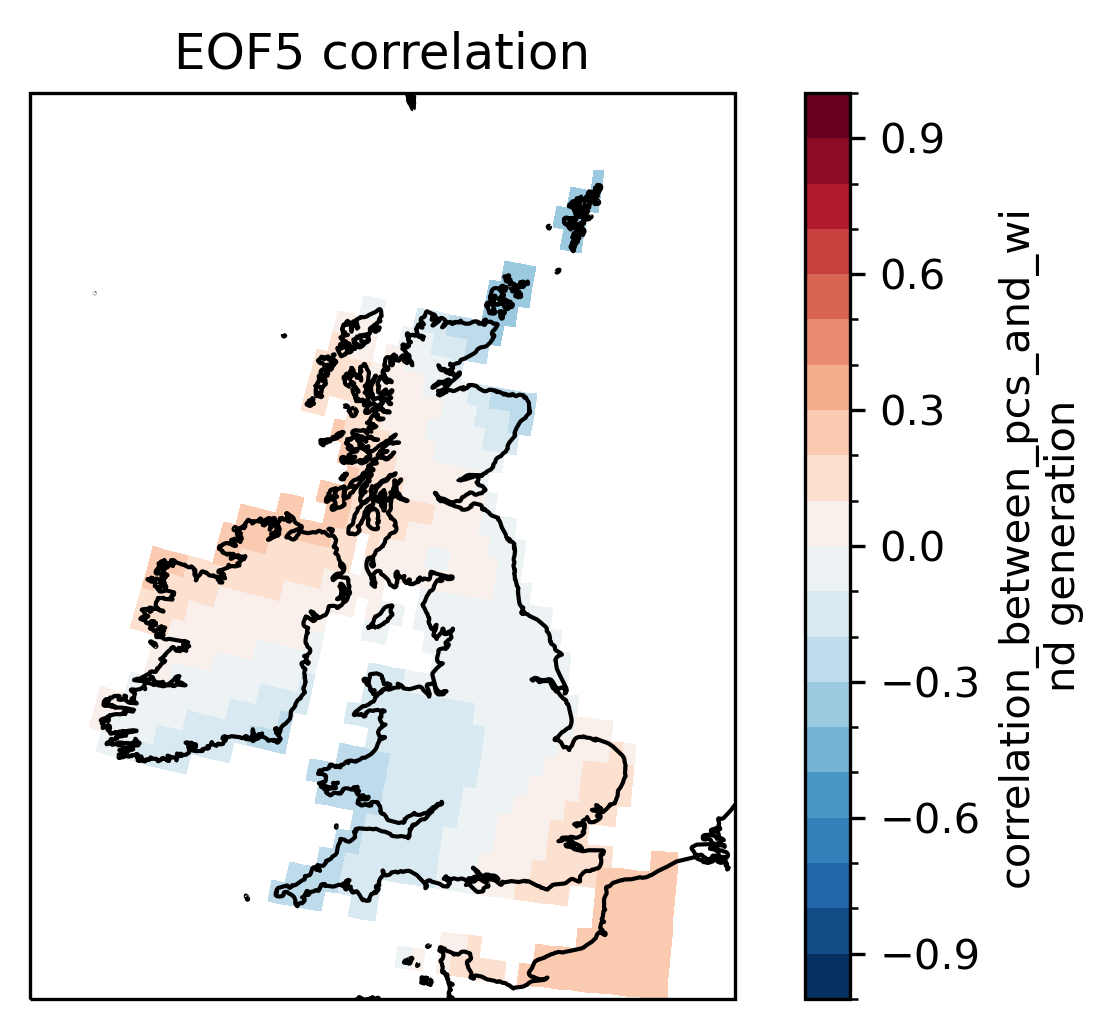}
             \caption{EOF 5}
             \label{fig:wind EOF5}
         \end{subfigure}
    \caption{The leading five EOFs for wind power in Britain, Ireland, and surrounding areas.}
    \label{fig:wind EOFs}
\end{figure}

The leading five EOFs for wind power in Britain, Ireland, and the surrounding areas are shown in Figure \ref{fig:wind EOFs}. Almost half of the variance in wind is captured in the first EOF, shown in Figure \ref{fig:wind EOF1}, which is highly and postively correlated with the entire area. North-south and east-west patterns account for the next highest shares in wind potential (Figures \ref{fig:wind EOF2} and \ref{fig:wind EOF3}), followed by coastal vs. inland patterns (Figures \ref{fig:wind EOF4} and \ref{fig:wind EOF5}).

\begin{figure}
    \centering
    \begin{subfigure}[b]{0.4\textwidth}
             \centering
             \includegraphics[width=\textwidth]{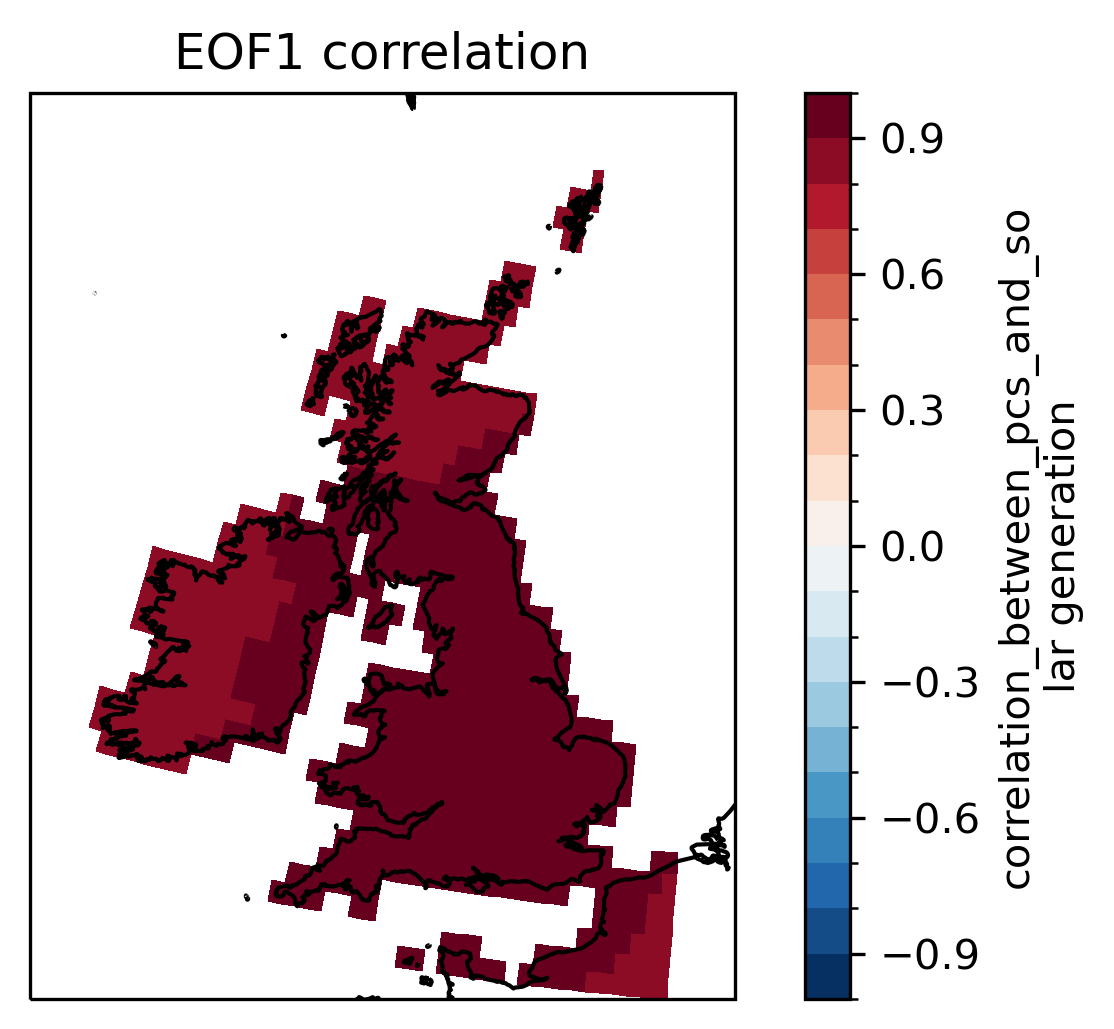}
             \caption{EOF 1}
             \label{fig:solar EOF1}
         \end{subfigure}
         \begin{subfigure}[b]{0.4\textwidth}
             \centering
             \includegraphics[width=\textwidth]{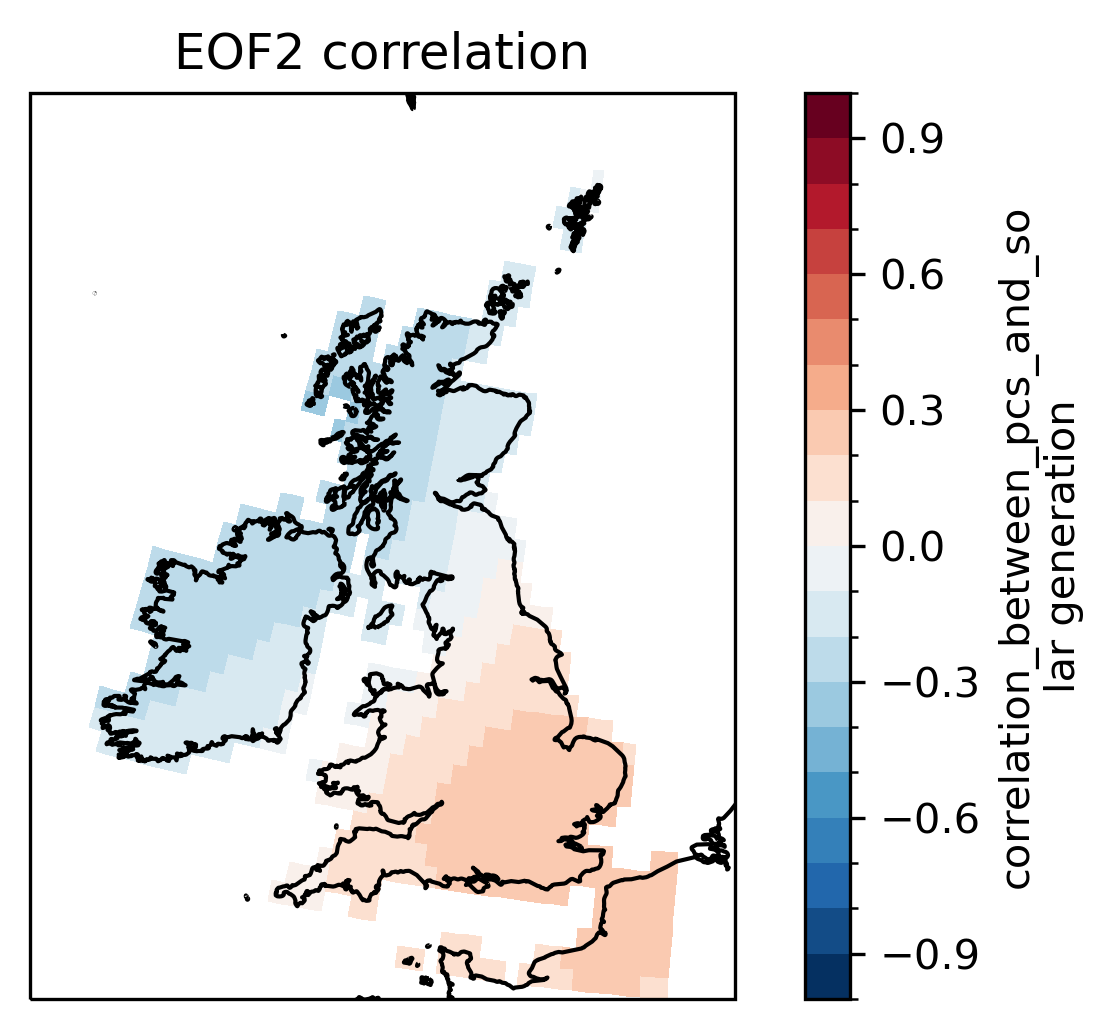}
             \caption{EOF 2}
             \label{fig:solar EOF2}
         \end{subfigure}
         \begin{subfigure}[b]{0.4\textwidth}
             \centering
             \includegraphics[width=\textwidth]{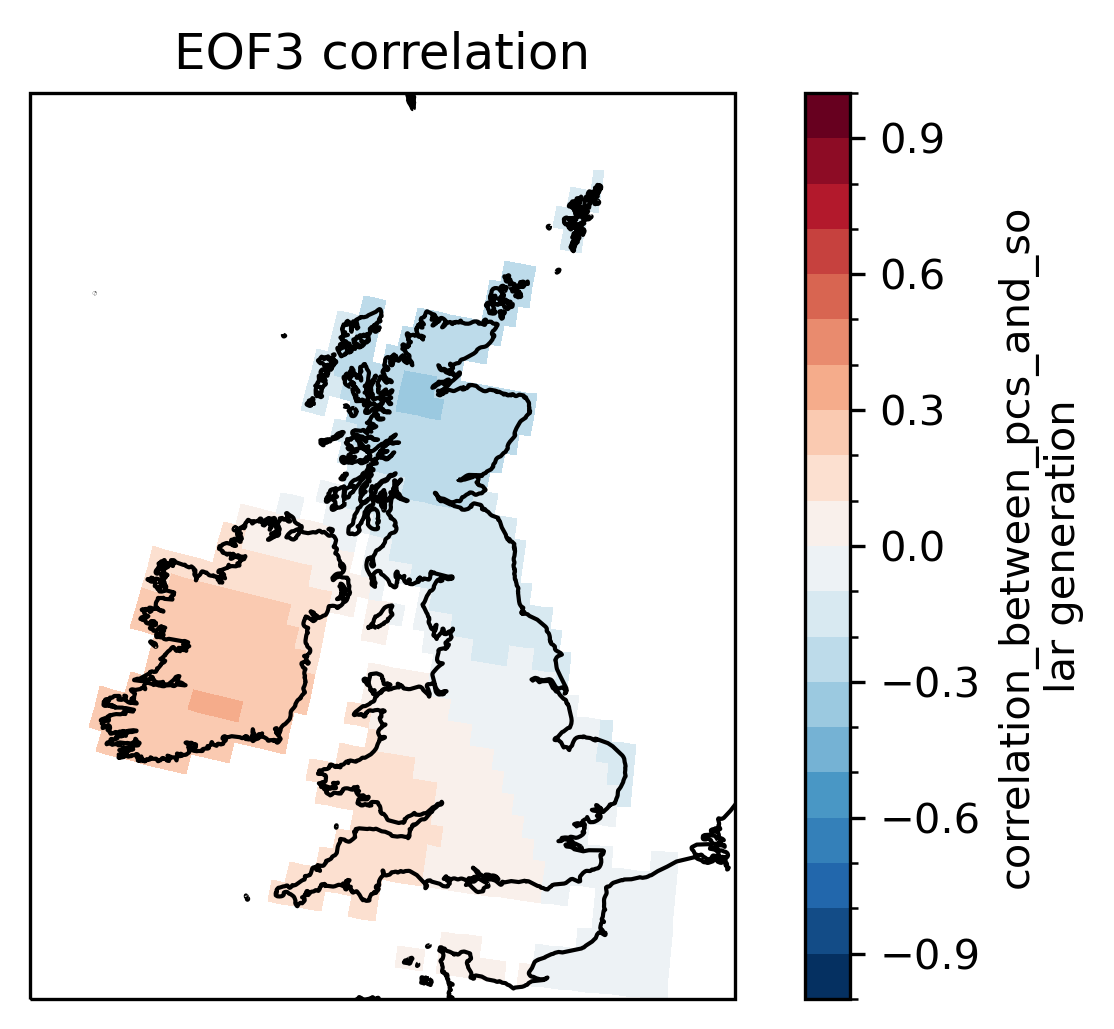}
             \caption{EOF 3}
             \label{fig:solar EOF3}
         \end{subfigure}
        \begin{subfigure}[b]{0.4\textwidth}
             \centering
             \includegraphics[width=\textwidth]{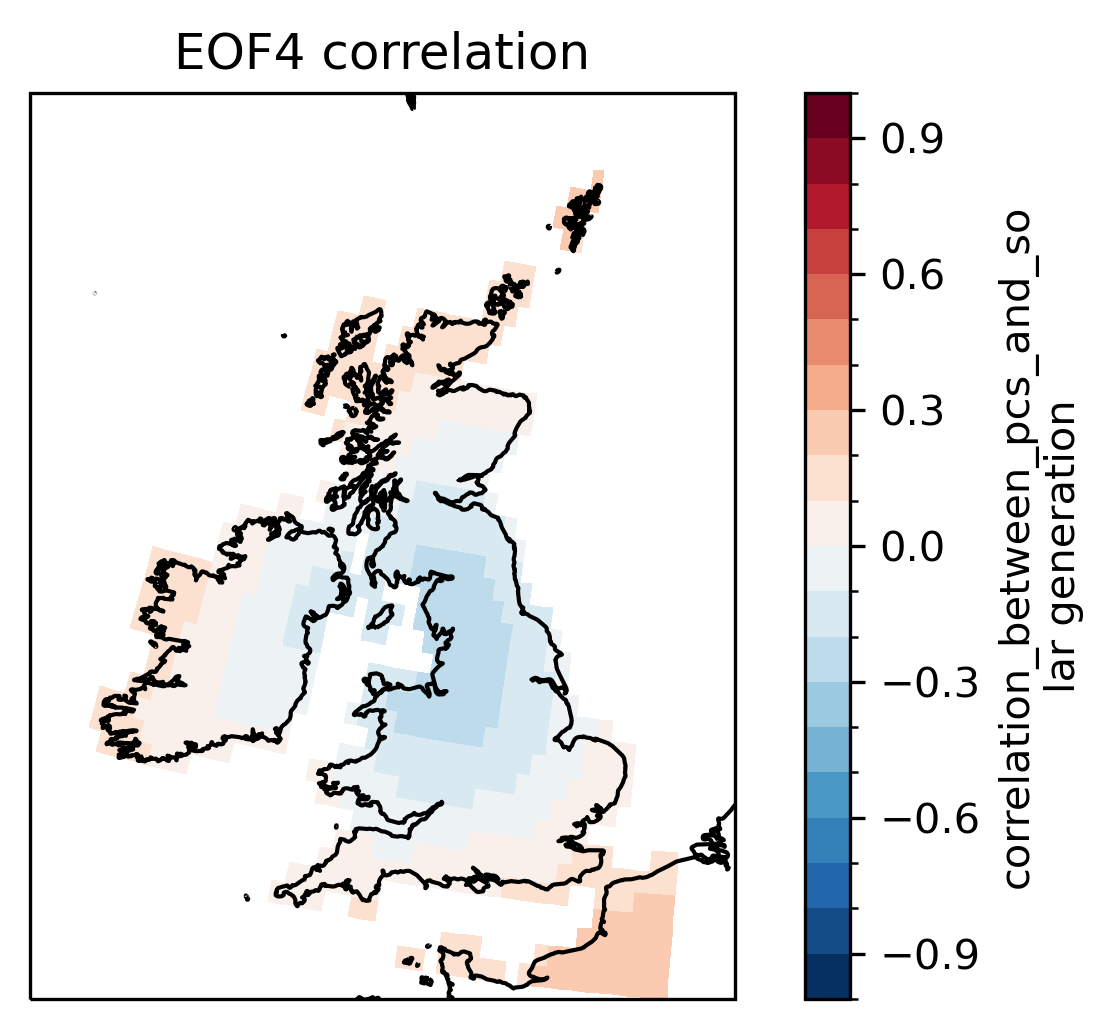}
             \caption{EOF 4}
             \label{fig:solar EOF4}
         \end{subfigure}
    \caption{The leading four EOFs for photovoltaic power in Britain, Ireland, and surrounding areas.}
    \label{fig:solar EOFs}
\end{figure}

The leading four EOFs for solar generation potential are shown in Figure \ref{fig:solar EOFs}. Over 80\% of solar variation is captured in the first EOF, shown in Figure \ref{fig:solar EOF1}. The corresponding PC captures seasonal and diurnal patterns as well as large storm systems that effect solar generation in the entire region. The following three EOFs capture northeast-southwest (Figure \ref{fig:solar EOF2}), east-west (Figure \ref{fig:solar EOF3}, and coastal-inland patterns (Figure \ref{fig:solar EOF4}).

\subsection{Max-p regionalization}

\begin{figure}
    \centering
         \begin{subfigure}[b]{0.35\textwidth}
             \centering
             \includegraphics[width=\textwidth]{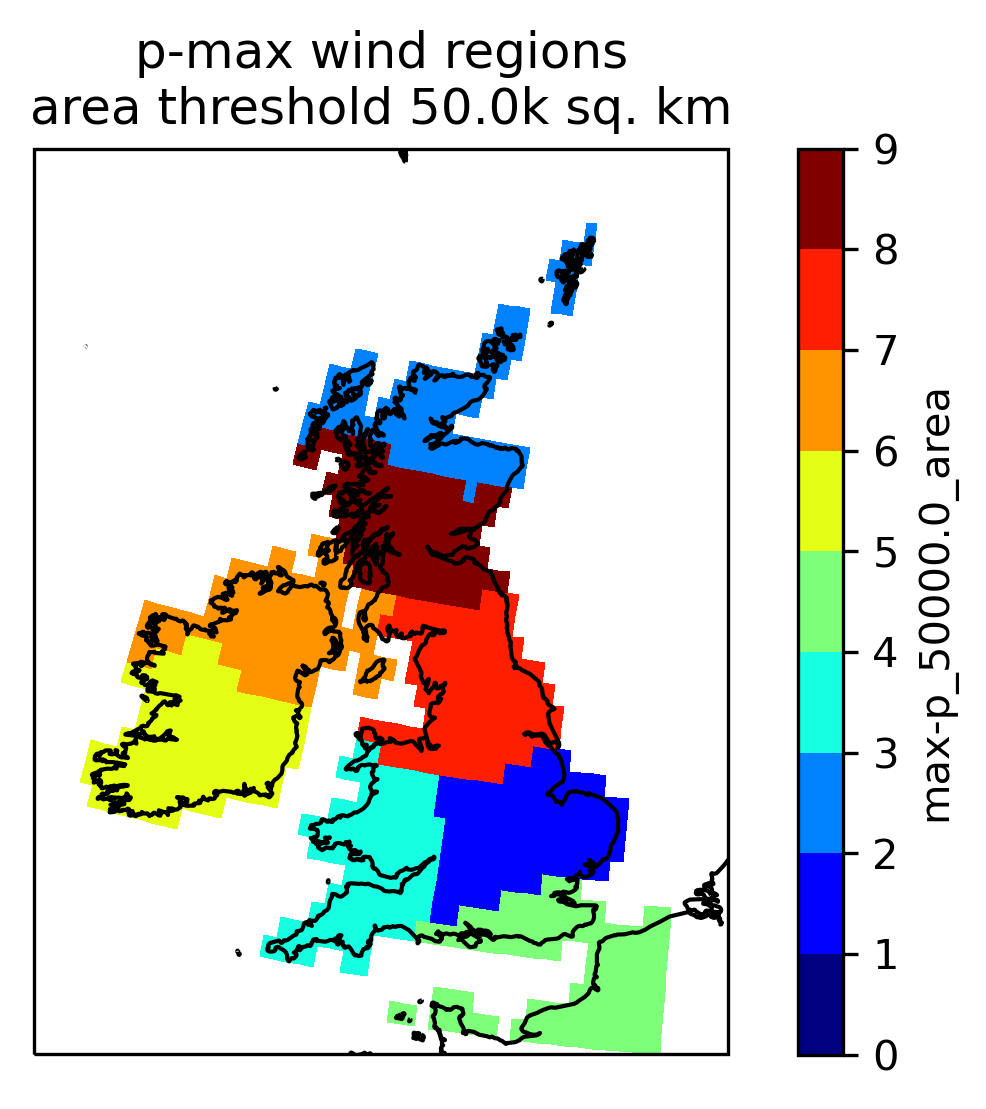}
            \caption{50,000 km$^2$ threshold regions}
             \label{fig:50k wind regions}
         \end{subfigure}
        \begin{subfigure}[b]{0.5\textwidth}
             \centering
             \includegraphics[width=\textwidth]{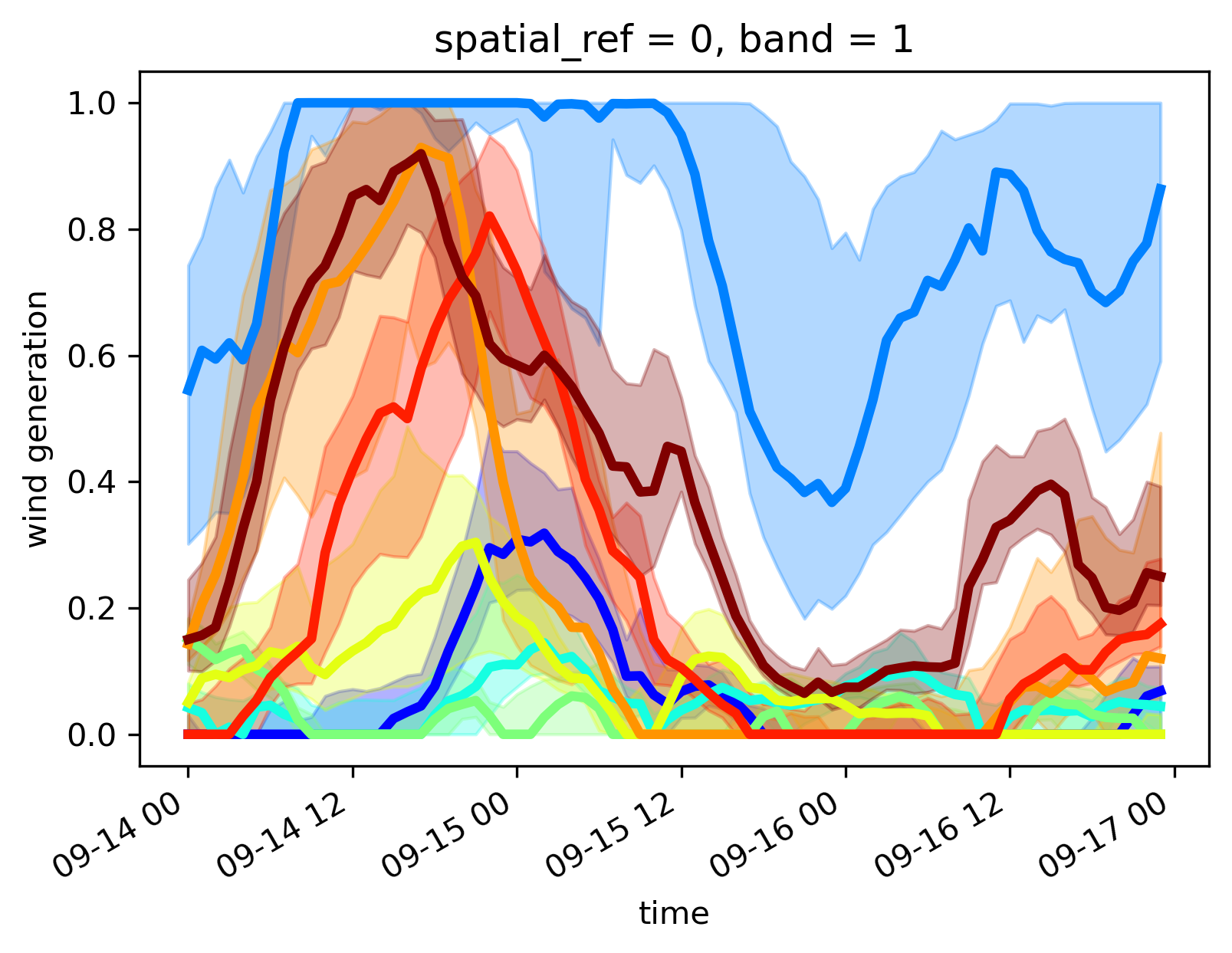}
            \caption{50,000 km$^2$ threshold generation, Silhouette: 0.153}
             \label{fig:50k wind pattern}
         \end{subfigure}
         \begin{subfigure}[b]{0.35\textwidth}
             \centering
             \includegraphics[width=\textwidth]{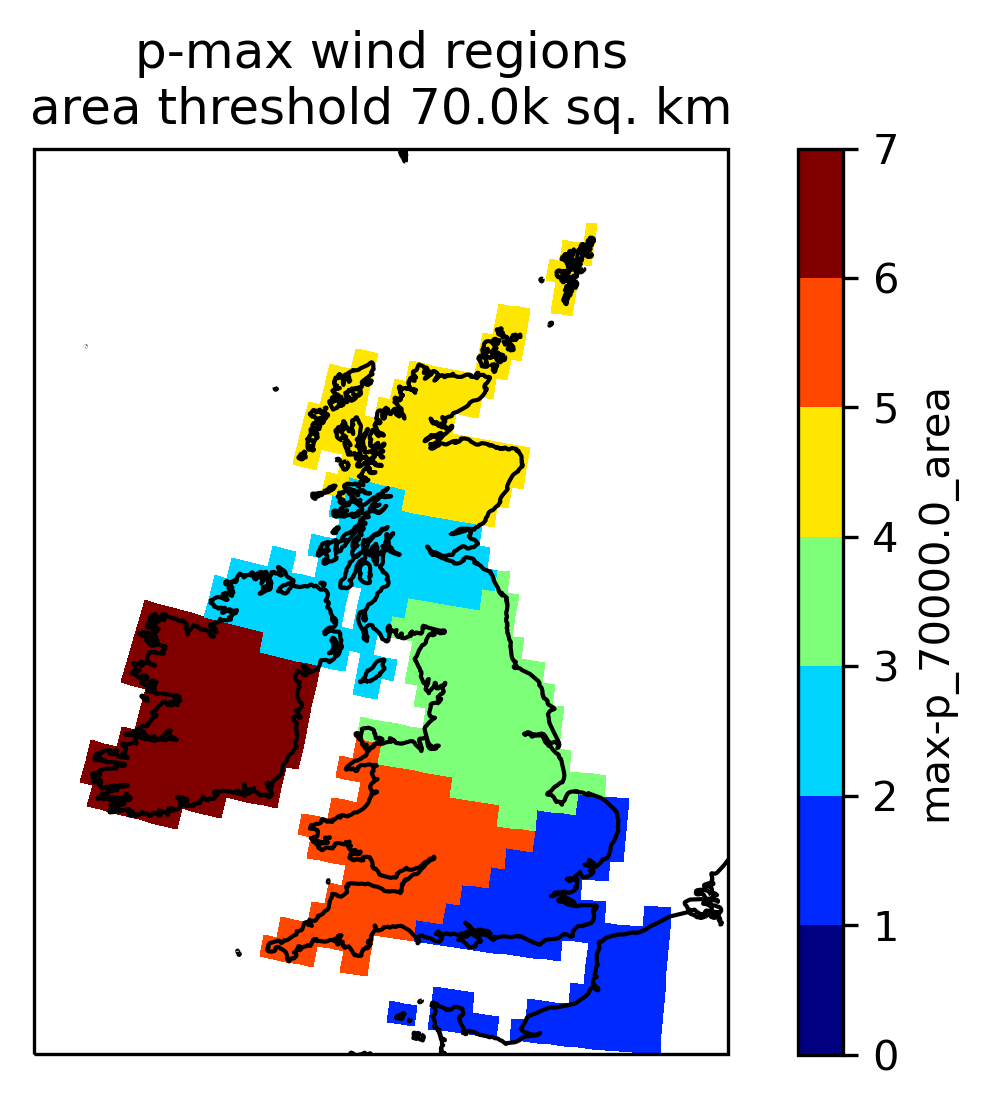}
            \caption{70,000 km$^2$ threshold regions}
             \label{fig:70k wind regions}
        \end{subfigure}
        \begin{subfigure}[b]{0.5\textwidth}
             \centering
             \includegraphics[width=\textwidth]{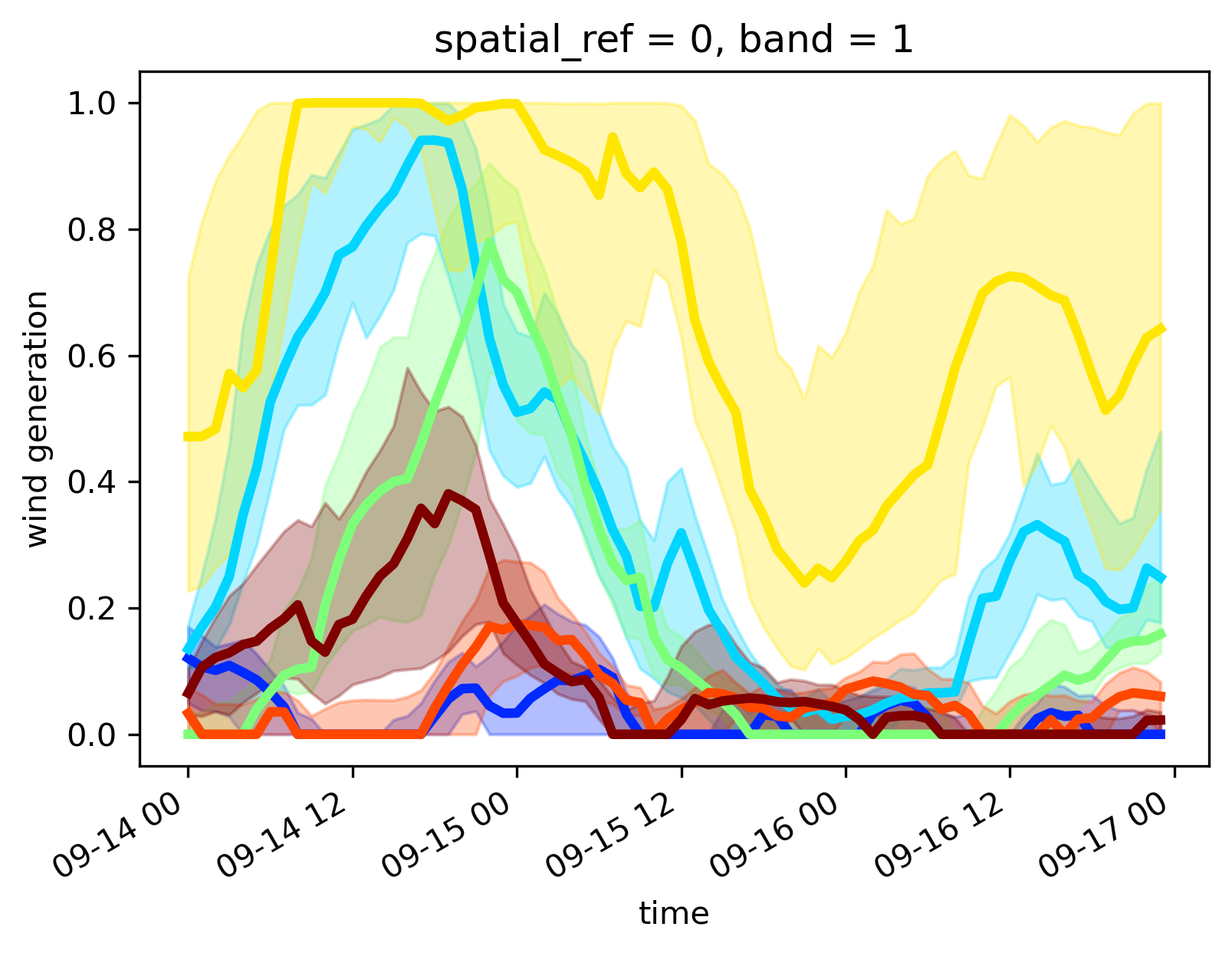}
            \caption{70,000 km$^2$ threshold generation, Silhouette: 0.175}
             \label{fig:70k wind pattern}
        \end{subfigure}
        \begin{subfigure}[b]{0.35\textwidth}
             \centering
             \includegraphics[width=\textwidth]{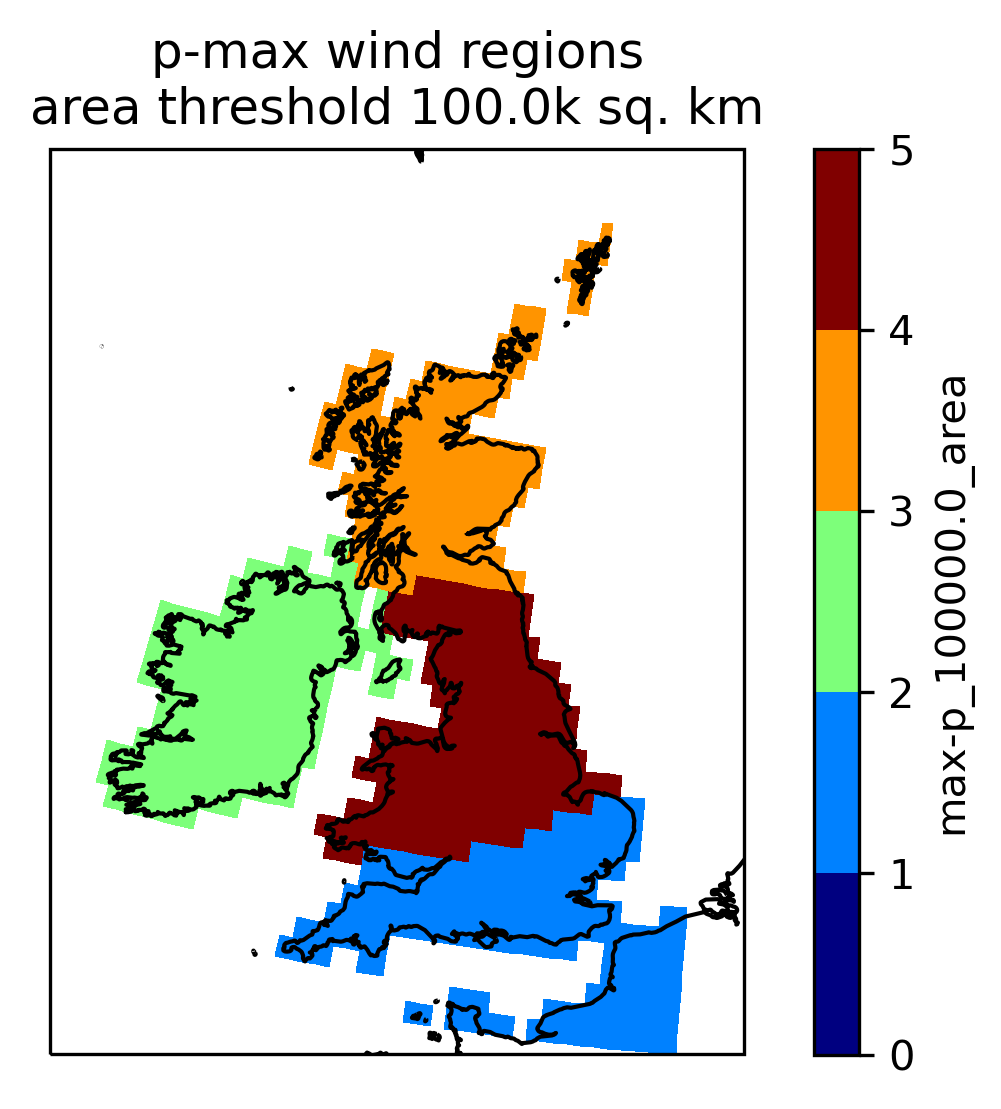}
             \caption{100,000 km$^2$ threshold regions}
             \label{fig:100k wind regions}
         \end{subfigure}
        \begin{subfigure}[b]{0.5\textwidth}
             \centering
             \includegraphics[width=\textwidth]{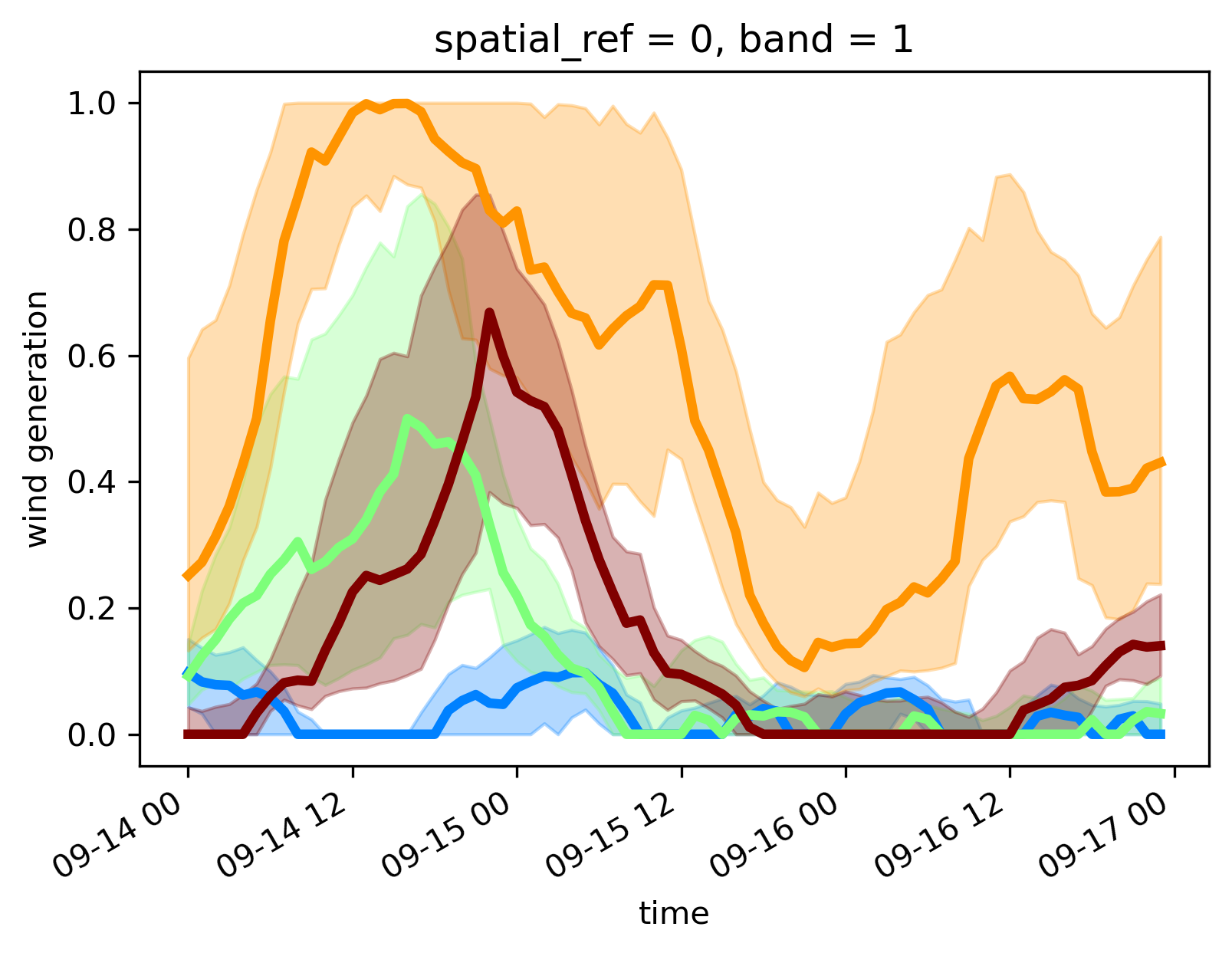}
             \caption{100,000 km$^2$ threshold generation, Silhouette: 0.179}
             \label{fig:100k wind patterns}
         \end{subfigure}
        \caption{Wind regions for area thresholds from 50,000 km$^2$ to 100,000 km$^2$ in Britain, Ireland, and surrounding areas and median, 25th, and 75th percentile generation profiles for 14 to 16 September.}
        \label{fig:wind maxp}
    \label{fig:enter-label}
\end{figure}

The clustered wind regions and sample generation patterns are shown in Figure \ref{fig:wind maxp}. Clustered regions are shown in Figures \ref{fig:50k wind regions}, \ref{fig:70k wind regions}, and \ref{fig:100k wind regions}. As the land area threshold increases, regional boundaries capture wind generation differences between Ireland and Britain, as well as north-south differences in Britain. 

The sample generation patterns in Figures \ref{fig:50k wind pattern}, \ref{fig:70k wind pattern}, and \ref{fig:100k wind patterns} display the median wind profile for each region for 14 to 16 September, with the range from the 25th to 75th percentile values filled. These profiles show that areas within a clustered region generally have similar temporal patterns in wind generation even as the range of generation values increases with larger threshold values. The Silhouette Cofficients range from 0.153 for the 50,000 km$^2$ land area threshold to 0.179 for the largest land area threshold of 100,000 km$^2$. These near-zero values indicate that there is overlap between wind generation potential in the regions, as expected for adjacent regions.

\begin{figure}[!htb]
    \centering
         \begin{subfigure}[b]{0.35\textwidth}
             \centering
             \includegraphics[width=\textwidth]{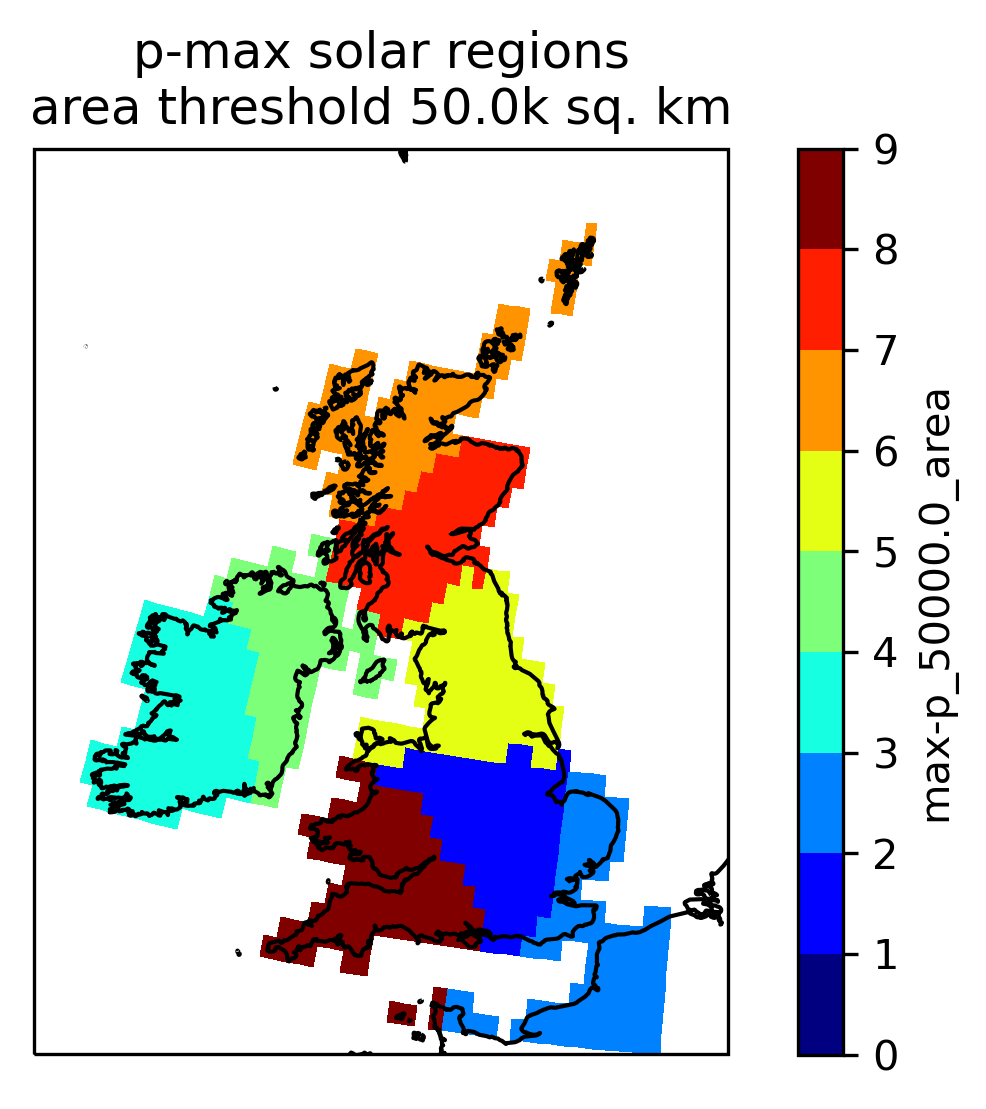}
            \caption{50,000 km$^2$ threshold regions}
             \label{fig:50k solar regions}
         \end{subfigure}
        \begin{subfigure}[b]{0.5\textwidth}
             \centering
             \includegraphics[width=\textwidth]{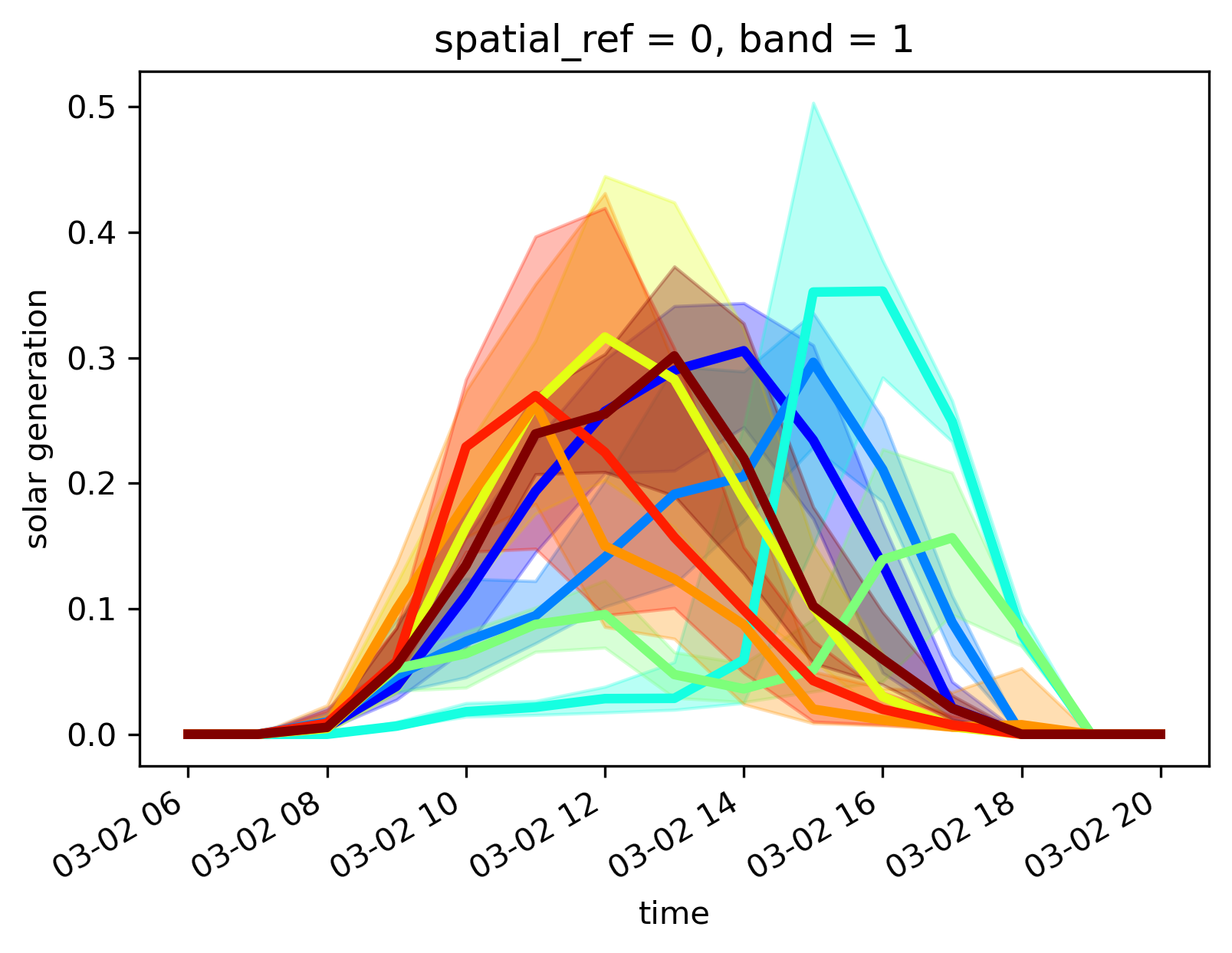}
            \caption{50,000 km$^2$ threshold generation, Silhouette: 0.125}
             \label{fig:50k solar patterns}
         \end{subfigure}
         \begin{subfigure}[b]{0.35\textwidth}
             \centering
             \includegraphics[width=\textwidth]{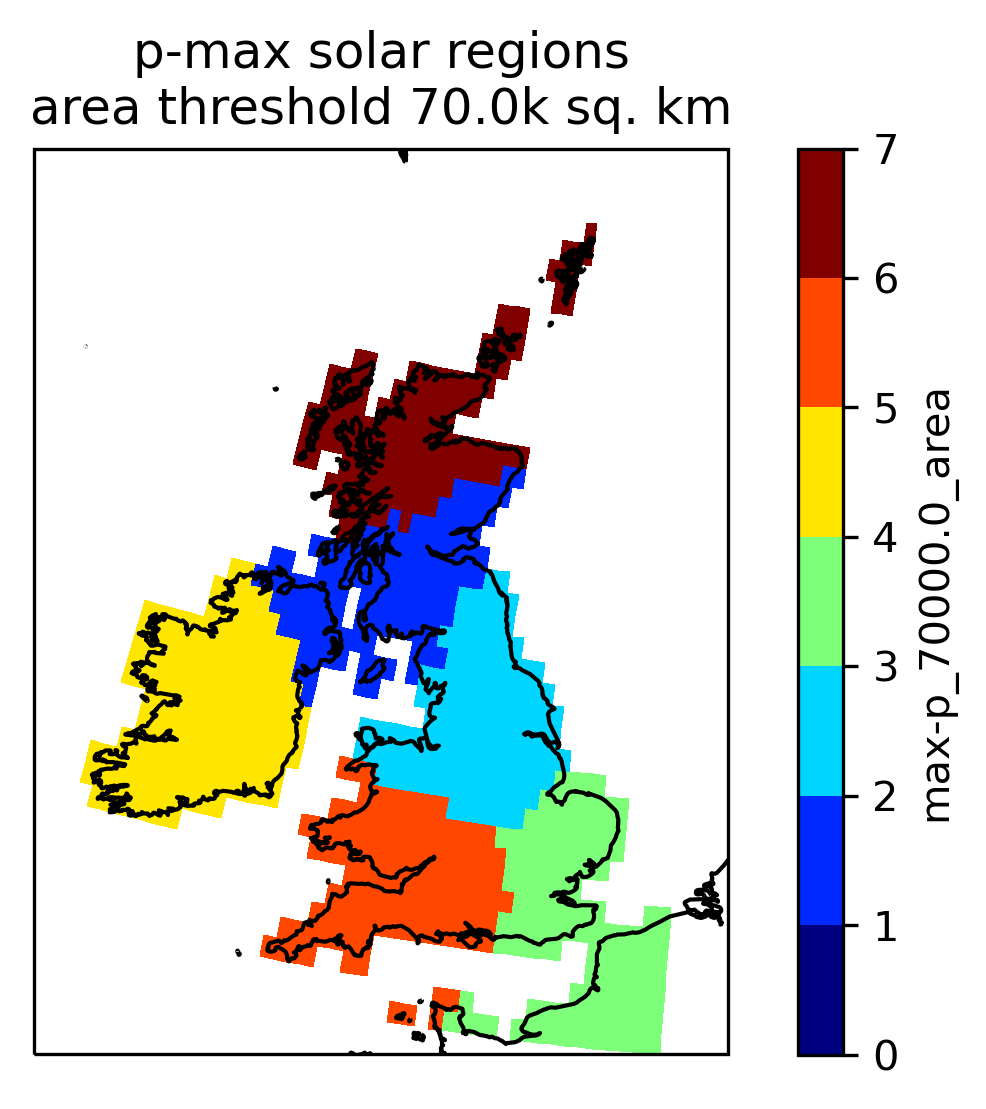}
            \caption{70,000 km$^2$ threshold regions}
             \label{fig:70k solar regions}
         \end{subfigure}
        \begin{subfigure}[b]{0.5\textwidth}
             \centering
             \includegraphics[width=\textwidth]{Figures/Solar_area_threshold_50.0k_sq._km_plot.png}
            \caption{70,000 km$^2$ threshold generation, Silhouette: 0.169}
             \label{fig:70k solar patterns}
        \end{subfigure}
        \begin{subfigure}[b]{0.35\textwidth}
             \centering
             \includegraphics[width=\textwidth]{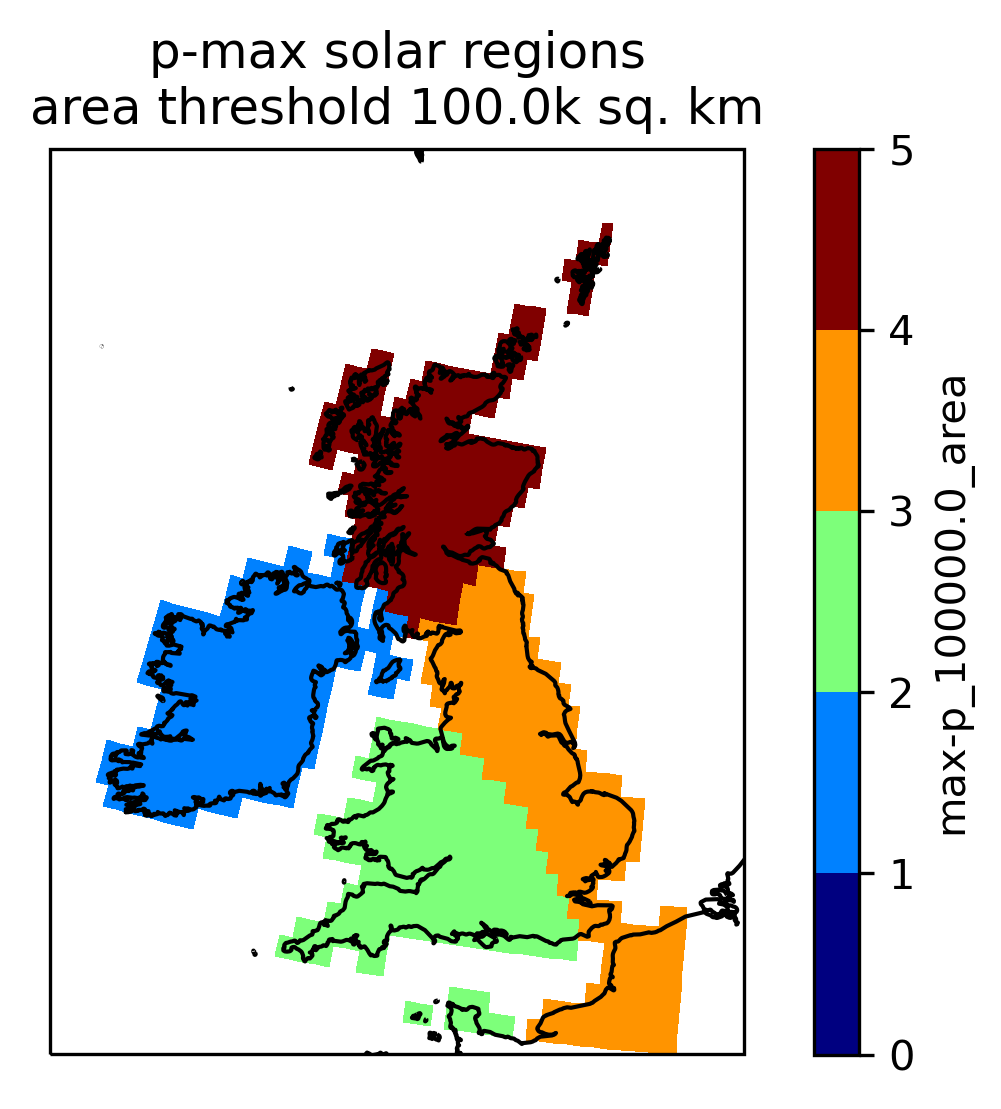}
            \caption{100,000 km$^2$ threshold regions}
             \label{fig:100k solar regions}
        \end{subfigure}
        \begin{subfigure}[b]{0.5\textwidth}
             \centering
             \includegraphics[width=\textwidth]{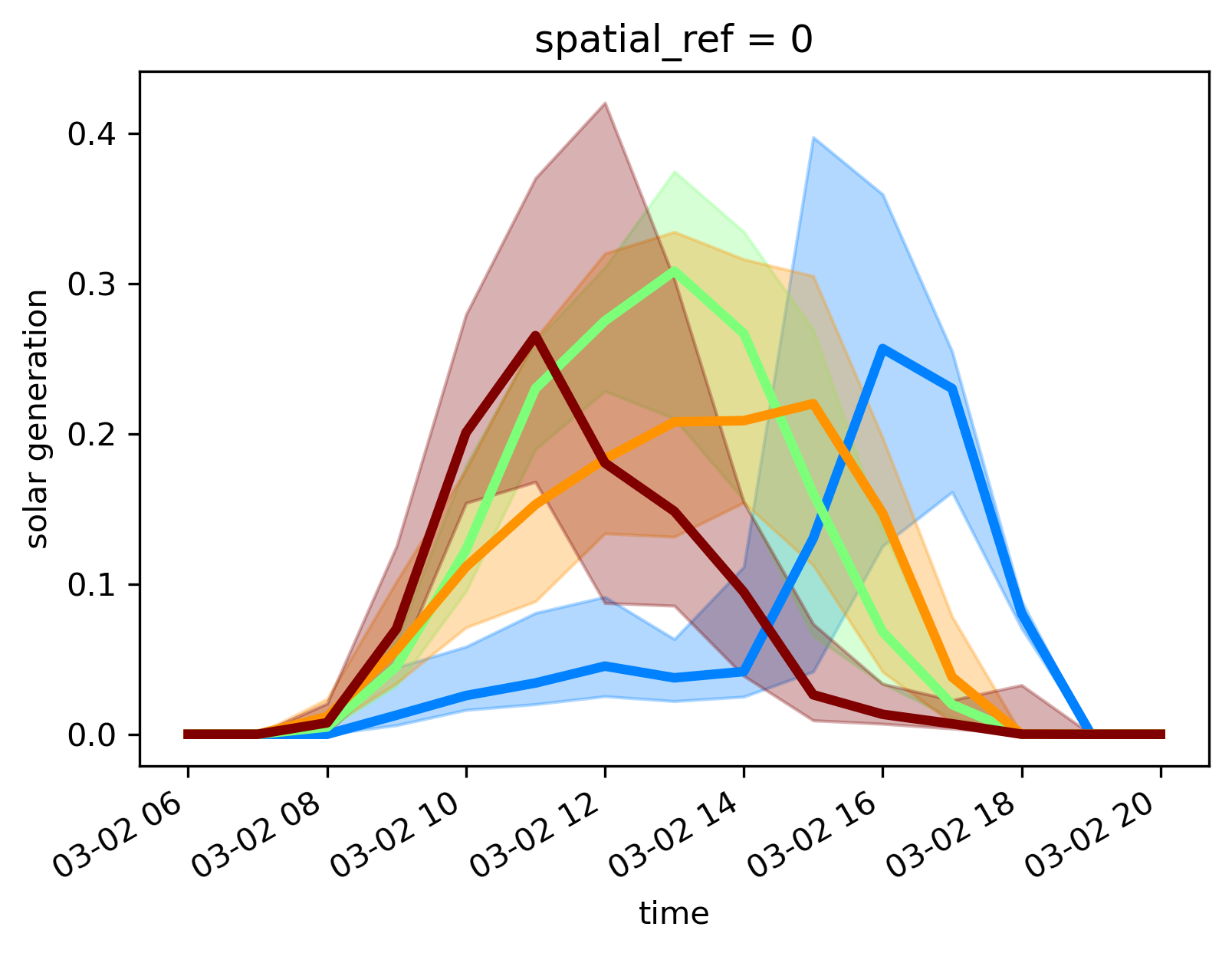}
            \caption{100,000 km$^2$ threshold regions, Silhouette: 0.154}
             \label{fig:100k solar patterns}
         \end{subfigure}
        \caption{Solar regions for area thresholds from 50,000 km$^2$ to 100,000 km$^2$ in Britain, Ireland, and surrounding areas and median, 25th, and 75th percentile generation profiles for 6:00 to 20:00 on 2 March.}
    \label{fig:solar maxp}
\end{figure}

The clustered solar regions and sample generation patterns are shown in Figure \ref{fig:solar maxp}. Compared to the wind regions in Figure \ref{fig:wind maxp}, these regions capture more east-west differences in solar potential.

Sample solar generation patterns for 06:00 to 20:00 on 2 March are shown in Figures \ref{fig:50k solar patterns}, \ref{fig:70k solar patterns}, and \ref{fig:100k solar patterns}. Although there are large ranges between the 25th and 75th percentile values for each time, areas within each cluster have similar temporal patterns in solar potential and tend to peak at similar times. The Silhouette Coefficients range from 0.125 for the 50,000 km$^2$ land area threshold to 0.169 for the 70,000 km$^2$ land area threshold. Like the wind generation profiles, these near-zero values indicate overlap between the solar profiles in each cluster. The largest Silhouette Coefficient occurs at a smaller spatial scale of 70,000 km$^2$, compared to the wind profiles, where the largest Silhouette Coefficient occurred at the largest land area threshold considered.

\section{Conclusion}
We introduce a method for identifying areas with similar temporal energy profiles by combining EOF analysis with max-p regionalization. We demonstrate this method for a case study of hourly wind and solar potential in 2019 in Britain and Ireland and show that this technique creates regions with similar wind and solar potential profiles. Of the land area thresholds considered, regions with a threshold of 100,000 km$^2$ had the best Silhouette Coefficient for the wind generation profiles and regions with a threshold of 70,000 km$^2$ had the best Silhouette Coefficient for solar generation profiles, indicating better-defined clusters occur at smaller spatial scales for solar generation compared to wind generation.

\section*{Acknowledgements}
Claire Halloran acknowledges the Rhodes Trust.
\bibliographystyle{ieeetr} 
\bibliography{references}
\end{document}